\documentstyle[preprint,revtex]{aps}
\begin{document}
\draft
\preprint{IFP-229-UNC}
\preprint{April, 1992}
\begin{title}
\begin{center}
$QCD$ renormalization for the top-quark mass\\
 in a mass geometrical mean hierarchy
\end{center}
\end{title}
\author{Daniel Ng}
\begin{instit}
\begin{center}
Institute of Field Physics\\
Department of Physics and Astronomy\\ University of North Carolina\\
Chapel Hill,  North Carolina 27599-3255
\end{center}
\end{instit}
\begin{abstract}
$QCD$ renormalization for the top-quark mass is calculated
in a mass geometrical
mean hierarchy, $m_d m_b = m_s^2$ and $m_u m_t = m_c^2$.  The physical mass,
$m_t(m_t) = 160 {\pm} 50 GeV$ is obtained, which agrees very well with
electroweak precision measurement.
\end{abstract}
\pacs{PACS numbers : 12.15Ji, 11.30Hv}
In the quark sector within the context of the standard model, there are 6
parameters for the quark masses and 3 for the mixing angles (if CP phase is
not included). From the empirical facts that
$V_{us} \simeq \sqrt{{m_d}/{m_s}}$
and $V_{us}^2 \simeq V_{cb} \simeq m_s/m_b$,
the mass of down quarks are related by $m_d m_b = m_s^2$.  Inspired by this
geometrical mean hierarchy, Ng and Ng \cite{ng91} suggested that the up quark
masses were also related by $m_u m_t = m_c^2$ and the third angle was given by
$V_{ub} \simeq {m_c}/{m_t}$.  In this case, there are only 4
parameters for the quark masses and the mixing angles are given by the
ratios of the quark masses.  Since there is large experimental uncertainty of
the measurement of $V_{ub}$\cite{review90}, therefore we use
the geometric mass relation to determine the top-quark mass.  Using
$m_u(1GeV) = 5.1 \pm 1.5 MeV$ and $m_c(1GeV) = 1.35 \pm 0.05 GeV$
\cite{gasser82},
we find that $m_t(1GeV) = 360 \pm 110 GeV$.  Within one standard deviation,
this
value still agrees with the electroweak precision measurement which predicts
that
%$m_t = 124 \begin{array}{c} \scriptstyle{+28+20}\\ \scriptstyle{-34-15}
%\end{array} GeV$
$m_t = 124 ^{+28+20}_{-34-35} GeV$
\cite{langacker91} and
$m_t = 140 \pm 35 GeV$ \cite{altarelli91}.

In fact, the masses are energy dependent.  In low energy, $QCD$ effect is
large and more important than the electroweak effect.  The two-loop
renormalization group equations for the strong coupling constant and
the quark mass are given by
\begin{equation}
\mu{\partial\over\partial \mu} \alpha = - \beta_0 {{\alpha^2}\over {2 \pi}}
- \beta_1 {{\alpha^3}\over {8 \pi^2}}\; ,
\end{equation}
and
\begin{equation}
\mu{\partial \over\partial \mu} m = - \bigg\lbrack \gamma_0 {\alpha\over \pi} +
 \gamma_1 ({\alpha\over \pi})^2 \bigg\rbrack m \; ,
\end{equation}
with $\beta_0=11-(2/3)N_f$, $\beta_1 = 102-(38/3)N_f$,
$\gamma_0 = 2$ and $\gamma_1 = 101/12 - (5/18)N_f$.
$N_f$ is the number of quark flavors and $\mu$ is the
energy scale.  The contribution from the electroweak sector is expected to be
smaller than that of $QCD$ in low energy regime, $\mu \leq m_t$.  In addition,
we use $N_f = 5$ for $ 1 GeV \leq \mu \leq m_t$.
Since all $\beta_0$, $\beta_1$, $\gamma_0$ and $\gamma_1$
are positive, the running mass,$m(\mu)$, decreases as $\mu$ increases.
Therefore, the physical mass of the top-quark,
$m_t(m_t)$, is expected to be substantially smaller
than $360 GeV$.

The geometrical mean relations are more precisely written as $m_d(\mu) m_b(\mu)
= \lbrack m_s(\mu) \rbrack ^2$ and $m_u(\mu) m_t(\mu)
= \lbrack m_c(\mu) \rbrack ^2$.  Since $QCD$ interaction is flavor diagonal,
therefore these relations hold for $\mu \leq m_t$.  In Ref.\cite{gasser82},
the authors obtained $m_d(1GeV) = 8.9 \pm 2.6 MeV$ , $m_s(1GeV) = 175 \pm
55 MeV$ and $m_b(1GeV)=5.3 \pm 0.1 GeV$ which follow the geometrical mean
hierarchy.  In the up quark sector, $m_u(1GeV) = 5.1 \pm 1.5 MeV$ and
$m_c(1GeV)
= 1.035 \pm 0.05 GeV$, where $\Lambda_{QCD} = 100 MeV$ and $N_f =3$ are used,
imply that $m_t(1GeV) = 360 \pm 110 GeV$.  For $\mu \geq 1 GeV$, we will use
$N_f$ = 5 in Eq.(1) and (2).  In the lowest approximation, $\alpha$ and $m_t$
in Eq. (1) and (2), which can be solved exactly, are given by
\begin{equation}
 \alpha(\mu) = {\alpha(1GeV)\over {1+\beta_0{{\alpha(1GeV)}\over {2\pi}}ln{\mu
 \over 1GeV}}}\; ,
\end{equation}
and
\begin{equation}
m_t(\mu)=m_t(1GeV) \bigg\lbrack 1+\beta_0{{\alpha(1GeV)}\over {2\pi}}ln{\mu
 \over 1GeV} \bigg\rbrack ^{-2 {\gamma_0\over \beta_0}}\; .
\end{equation}
{}From the global average \cite{opal92}, $\alpha(M_Z) = 0.122 \pm 0.006$, we
obtain $\alpha(1GeV) = 0.37 \pm 0.06$.  Therefore, $m_t = 190 \pm 60 GeV$
is obtained.  On the other hand, Eq.(1) and (2), keeping $\beta_1$ and
$\gamma_1$ terms, can be solved numerically.  The result for $m_t(m_t)$ is
\begin{equation}
 m_t(m_t) = 160 \pm 50 GeV
\end{equation}
which agrees very well with the mass range predicted from electroweak
precision measurement \cite{langacker91,altarelli91}.

In this note, we have calculated the $QCD$ renormalization for the top-quark
mass predicted by the mass geometrical mean hierarchy, namely,
$m_d m_b = m_s^2$ and $m_u m_t = m_c^2$.
We obtain $m_t = 160 \pm 50 GeV$ which agrees
very well with the prediction from the electroweak precision measurement.
The calculation was based on the fact that the geometrical mean hierarchy
holds for $\mu \leq m_t$ since $QCD$ interaction is flavor diagonal and is
much larger than that of electroweak interaction.  New physics which enforces
this geometrical mean hierarchy will show up when the energy is above $m_t$.
In the lepton sector, the lepton masses , which do not follow the geometrical
mean hierarchy, however are related by $m_em_{\tau} \simeq (1/9) m_{\mu}^2$
which can be achieved in an unified theory \cite{georgi79}.  With this
interesting mass hierarchy, we require only 4 parameters, instead of 9, in the
quark sector.  Discovering a top-quark with mass ranging from $110 GeV$ to
$210 GeV$ will be an indication for this proposed mass geometrical mean
hierarchy.  A more reliable calculation for $m_u$ is necessary in order to
obtain a more definite prediction of the top-quark mass.

\acknowledgments
The author would like to thank K. S. Babu and Y. J. Ng for discussion.  This
work was supported by the U.S. Department of Energy under Grant No.
DE-FG05-95ER-40219.

\end{document}